\documentclass[a4paper,aps,onecolumn,showkeys,nofootinbib,preprintnumbers,superscriptaddress,amsmath,amssymb,amsfonts]{revtex4}
\usepackage{graphicx}
\usepackage{bm,natbib}
\usepackage{amsmath}
\usepackage[english]{babel}
\usepackage[T1]{fontenc}
\usepackage{amssymb}
\usepackage{amsfonts}
\usepackage{epsfig}
\usepackage{colordvi}
\usepackage{psfrag}
\usepackage{color}
\usepackage{ mathrsfs }
\newcommand{\Lagr}{\mathcal{L}}

\newcommand{\Gi}{\mathcal{G}}

\newcommand{\E}{\mathcal{E}}
\newcommand{\Sw}{\mathscr{S}}
\usepackage{dcolumn}
\usepackage{multirow}
\usepackage{hyperref}
\usepackage{epstopdf}
\usepackage{bm}
\usepackage{times}
\usepackage{comment}

\hypersetup{
  colorlinks=true,        
  linkcolor=blue,         
  citecolor=magenta,      
}

\begin{document}
\title{Renormalizability of alternative theories of gravity: differences between power counting and entropy argument}

\author{Francesco Bajardi}
\email{francesco.bajardi@unina.it}
\affiliation{Dipartimento di Fisica  ``E. Pancini", Universit\`a di Napoli  ``Federico II", Via Cinthia, I-80126, Napoli, Italy.}
\affiliation{Istituto Nazionale di Fisica Nucleare (INFN), sez. di Napoli, Via Cinthia 9, I-80126 Napoli, Italy.}

\author{Francesco Bascone}
\email{francesco.bascone@unina.it}
\affiliation{Dipartimento di Fisica  ``E. Pancini", Universit\`a di Napoli  ``Federico II", Via Cinthia, I-80126, Napoli, Italy.}
\affiliation{Istituto Nazionale di Fisica Nucleare (INFN), sez. di Napoli, Via Cinthia 9, I-80126 Napoli, Italy.}

\author{Salvatore Capozziello}
\email{capozziello@unina.it}
\affiliation{Dipartimento di Fisica  ``E. Pancini", Universit\`a di Napoli  ``Federico II", Via Cinthia, I-80126, Napoli, Italy.}
\affiliation{Istituto Nazionale di Fisica Nucleare (INFN), sez. di Napoli, Via Cinthia 9, I-80126 Napoli, Italy.}
\affiliation{Scuola Superiore Meridionale, Largo San Marcellino 10, 80138 Napoli, Italy.}
\affiliation{Laboratory for Theoretical Cosmology, Tomsk State University of Control Systems and Radioelectronics (TUSUR), 634050 Tomsk,  Russia. }

\date{\today}


\begin{abstract}
It is well known that General Relativity cannot be considered under the standard of  a perturbatively renormalizable quantum field theory, but asymptotic safety is  taken into account as a possibility for  the formulation of gravity as a non-perturbative renormalizable theory. Recently, the entropy argument has however stepped into the discussion claiming for a "no-go" to the asymptotic safety argument. In this paper, we present simple counter-examples, considering alternative theories of gravity, to the entropy argument as further indications, among others, on the possible flows in the assumptions on which the latter is based on. We consider  different theories, namely curvature based extensions of General Relativity as  $f(R)$, $f(\Gi)$,  extensions of Teleparallel Gravity as  $f(T)$, and Horava-Lifshitz gravity, working out the explicit spherically symmetric solutions  in order to make a comparison between power counting and the entropy argument. Even in these  cases, inconsistencies are found.
\end{abstract}
\maketitle

\section{Introduction}
One of the main issues  arising when considering a quantum theory of gravity is the non-renormalizability of General Relativity (GR). More precisely, GR cannot be formulated as a perturbatively renormalizable quantum field theory. This can be  easily understood by  simple power counting arguments. However, starting from Weinberg results \cite{weinb}, researchers started looking for non-perturbative renormalizable quantum field  formulations of GR. This latter case is usually referred to as asymptotic safety (see \cite{Niedermaier2007,Reuter2007,Bonanno1,Bonanno2} for reviews on the subject). This belief relies on the possibility that there may be another non-trivial UV fixed point in the renormalization group flow of the gravitational field with respect to which GR can be considered as a perturbation by some relevant operator. In this case, the non-renormalizability of gravity would just mean that we are expanding in the wrong variables, and power counting fails for this reason. However, fixed point of asymptotic safety scenario are gauge-dependent; this issue can be addressed by considering a generalization of the standard effective action, the so-called gauge-fixing independent effective action. Although it formally yields the same S-matrix as the standard one, it is independent of the choice of the gauge condition \cite{Falkenberg:1996bq}. 

Starting from holographic aspects of gravity, seen as incompatible with the aim of turning GR into a local quantum field theory, one may be led to ask whether there could be some no-go result obstructing the asymptotic safety scenario. In \cite{Shomer:2007vq} (but the idea behind was present earlier in \cite{Aharony1999}) for example, the so-called entropy argument is adopted as an explicit argument against the GR renormalizability  in any spacetime dimension. The crux of the argument relies on the assumption that the high-energy spectrum of GR is dominated by black holes, in the sense that once enough energy is concentrated in a spacetime region, a black hole will form, claiming that the high energy behavior of any $n$-dimensional renormalizable quantum field theory should be that of a conformal field theory (CFT) in the same number of dimensions. Technically this means that the entropy of a generic conformal field theory and the entropy given by the Bekenstein-Hawking formula should have the same energy-scaling behavior.

Even though this argument appears reasonable from different points of view, its validity is still under debate. In fact, there are several problematic physical assumptions on which the argument is based on. The clearest expositions of the issues and the debate under discussion to our current knowledge is the one given in \cite{Doboszewski2018}. The main assumption of the entropy argument is the black hole dominance. This is surely reasonable from our low-energy expectation, but it is indeed borrowed from the low-energy approximation, so that it is legitimate wondering to what extent this expectation should be trusted. In fact, at the Planck scale, physics could be different from what we expect and, furthermore, our understanding of micro black holes is still dominated by conjectures. Shomer  \cite{Shomer:2007vq} pointed out that this counter-argument should not hold since asymptotic safety scenario is based on the assumption that gravity is a valid low energy approximation of some quantum field theory. Therefore, arguing against the non-renormalizability of gravity is like arguing against the validity of the Bekenstein-Hawking formula. However, this is another source of concern. Indeed, there is no \emph{a priori} reason to expect Bekenstein-Hawking formula to hold at  Planck scales, and, even more, if one considers the recent literature concerned about quantum corrections to the formula \cite{Sen2013, Bytsenko2013, Shankaranarayanan2012, Vagenas2002}.

Specifically, the statement that the $n$-dimensional Bekenstein-Hawking entropy is asymptotically the same as that in   $n$-dimensional CFT, in the high energy regime, suffers some inconsistencies. Indeed, it is basically due to the fact that, if we have a field of mass $M$ and take the energy to be $\E \gg M$, almost all of the energy will be momentum and the mass will be negligible, resulting in a  UV CFT fixed point. However, because of this, it is more appropriate to consider the entropy argument only as a necessary condition. This latter consideration make it only useful as a way to understand if a QFT is non-renormalizable.

What is usually done in GR is  to consider the energy-scaling behavior of a $n$-dimensional CFT as  $\Sw \sim \E^{\frac{n-1}{n}}$ and compare it with the energy-scaling of the entropy of a Schwarzschild black hole (considering, for instance, asymptotically flat spacetimes) via the Bekenstein-Hawking formula for the entropy:
\begin{equation}
{\Sw}=\frac{A}{4G} \sim \frac{r^{n-2}_H}{G} \sim \E^{\frac{n-2}{n-3}},
\end{equation}
with $G$ being the Newton constant,  $A$ the black hole surface area, and $r_H$ the horizon radius. In this case, it is evident that there is no dimension where the two formulas agree. According to this result,  it follows  that $n$-dimensional GR is non-renormalizable. The same can be done for negative cosmological constant, where the usual AdS/CFT correspondence in recovered as a manifestation of Holography. On the other hand, the positive cosmological constant (de Sitter) is instead peculiar and cannot be addressed in this way. 

In this paper,  we consider  counterexamples to this approach, analyzing four different modified theories of gravity and checking their coupling constant mass dimension via power counting (around a Gaussian fixed point, the superficial degree of divergence must give the correct information about the renormalizability of the quantum field theory, the other way around is prevented by the possibility that cancellations occur). Then we compare the results with the entropy argument, showing that the approach manifests a clear non-agreement in the comparison. Interestingly, we also show that spherically symmetric solutions of modified theories (e.g. $f(R) = R^k$ and $f(\Gi) = \Gi^k$) occur when the power-law parameter $k$ is linked to the dimension such that the theory results power counting renormalizable. 

Throughout the paper, we will use natural units: $\hbar = c = k_B = e^+ = 1$.

\section{A summary on  Extended and Modified  Theories of Gravity}
As a general remark, we are dealing with {\it Modified Gravity} \cite{Cai} when some basic assumption of GR, as Equivalence Principle, metric connection or Lorentz invariance are not requested at the foundation of the theory. On the other hand, {\it Extended Gravity} \cite{Rep} means that GR is recovered in some limit or as a particular choice in the action. To give an example, Teleparallel Equivalent General Relativity (TEGR), based on the torsion scalar $T$,  is a modified theory of gravity while $f(R)$ gravity, based on the curvature scalar $R$, is a straightforward   extension of GR.

This section aims to outline the main properties of some modified/extended theories of gravity, in order to deal with their spherically symmetric solutions and to compute power counting and Bekenstein-Hawking entropy in view of a possible renormalizability of gravitational interaction.  We mainly analyze four different  theories, namely $f(R)$, $f(\Gi)$, $f(T)$ and Horava-Lifshitz gravity. The idea of extending GR started up due to the need of explaining many inconsistencies at astrophysical and  cosmological level, but also for better adapting gravity with other fundamental interactions. With regards to the large scale, most modified/extended theories explain the accelerating expansion of universe by introducing new curvature or torsion contributions, instead of Dark Matter or Dark Energy \cite{Bahamonde:2018miw, Sharif:2012ig, Bamba:2012cp, Capozziello:2012ie, Nojiri:2006ri, Bajardi:2020fxh}. These geometric terms are also considered to for the early universe inflationary paradigm \cite{Nojiri:2007uq, Zhong:2018tqn, DeLaurentis:2017ezn, Ferraro:2006jd}. The simplest case is  $f(R)$ gravity, since it represents a natural extension of Einstein's GR to a generic function of the Ricci scalar curvature $R$, not required to be linear in the Einstein-Hilbert action. By relaxing the hypothesis of second-order field equations, it is possible to construct an action of the form $S = \int \sqrt{-g} f(R) \; d^4x$. By varying  with respect to the metric, one gets the following field equations:
\begin{equation}
f_R(R) R_{\mu \nu} - \frac{1}{2} g_{\mu \nu} f(R) + (g_{\mu \nu} \Box - \nabla_\mu \nabla_\nu)f_R(R)=0.
\end{equation}
The presence of higher-order terms yields several interesting aspects as well as massive gravitational waves \cite{Calmet:2017rxl, Capozziello:2015nga, Capozziello:2019klx} or modifications of the Newtonian potential \cite{DeLaurentis:2013ska, Bertolami:2007gv, Capozziello:2021goa}. Other theories leading to higher than fourth-order field equations can be constructed by the general action
\begin{equation}
S = \int \sqrt{-g} f(R, \Box R, \Box^2 R, ... \Box^k R) \; d^4x\,,
\end{equation}
where $\Box$ is the d'Alembert operator.
A detailed treatment can be found in \cite{Tomboulis:2015esa, Sanyal:2003tv, Capozziello:2012hm}. Another possibility is to deal with other higher-order curvature invariants, such as $R^{\mu \nu} R_{\mu \nu}, \, R^{\mu \nu \rho \sigma} R_{\mu \nu \rho \sigma}$. Among possible extensions involving these second-order scalars, a particular combination gives rise to the topological surface term $\Gi$, defined as:
\begin{equation}
\Gi \equiv R^2 - 4 R^{\mu \nu} R_{\mu \nu} +  R^{\mu \nu \rho \sigma} R_{\mu \nu \rho \sigma},
\label{GB scalar}
\end{equation}
the so-called \emph{Gauss-Bonnet} topological invariant. In four dimensions, it represents the Pfaffian of the curvature form of the Levi-Civita connection, and, due to the generalized Gauss-Bonnet theorem, its integral over the manifold provides the Euler Characteristic of the manifold. For this reason, in four dimensions, it does not provide any contribution to the dynamics. Nevertheless, a function of $\Gi$ is topologically trivial only in less than 3+1 dimensions and can be used into the gravitational action in order to study the dynamics of the system. By varying $f(\Gi)$ with respect to the metric leads to the  field equations:
\begin{align}\label{feq}
\frac{1}{2} g_{\mu \nu} f(\Gi) &- \left(2R R_{\mu \nu} - 4 R_{\mu \alpha} R^{\alpha}{}_\nu + 2 R_\mu {}^{\alpha \beta \gamma} R_{\nu \alpha \beta \gamma} - 4 R^{\alpha \beta} R_{\mu \alpha \nu \beta}\right) f_\Gi(\Gi) +\nonumber
\\
&+ \left[2R \nabla_\mu \nabla_\nu +4 G_{\mu \nu} \Box - 4 (R_{ \nu}^\rho \nabla_{\mu} +R_{ \mu}^\rho \nabla_{\nu}) \nabla_\rho + 4 g_{\mu \nu} R^{\rho \sigma} \nabla_\rho \nabla_\sigma - 4 R_{\mu \alpha \nu \beta} \nabla^\alpha \nabla^\beta \right] f_\Gi(\Gi) = 0.
\end{align}
An important advantage of this formulation of gravity is given by the absence of ghost modes in  gravitational waves, as discussed in \cite{Capozziello:2019klx}; even in cosmology the theory yields many interesting predictions which alleviate the GR  issues at infrared scales \cite{baj&capo, Elizalde:2018qbm, Benetti:2018zhv, Paul:1990aa, Bajardi:2020mdp}. 

Another theory we are going to consider is TEGR and its extensions.  It is mainly aimed at solving  problems occurring in GR by treating gravity as a gauge theory of the translation group in the flat tangent spacetime. 
The flat and curved spacetime have a point by point correspondence which is mathematically represented by tetrad field through the relation:
\begin{equation}
g_{\mu \nu} e^\mu_a e^\nu_b = \eta_{ab},
\end{equation}
where the Latin indices label the flat spacetime, while the Greek indices label the curved spacetime. The fundamental field is represented by  torsion, while  curvature vanishes identically after neglecting the spin connections. The action reads as:
\begin{equation}
S= \frac{1}{16 \pi G} \int d^4x \, |e| \, T,
\label{tegr action}
\end{equation} 
where $e$ is the determinant of tetrad fields $e^a_{\mu}$, and $T$ is the torsion scalar, defined as 
\begin{equation}
T=S^{\rho \mu \nu} \, T_{\rho \mu \nu},
\end{equation}
with $S^{\rho \mu \nu}$ being the so called "Superpotential"
\begin{equation}
\label{superpot}
S^{\rho \mu \nu}=K^{\mu \nu \rho}-g^{\rho \nu} T_{\,\,\,\,\,\,\, \sigma}^{\sigma \mu}+g^{\rho \mu} T_{\,\,\,\,\,\,\, \sigma}^{\sigma \nu},
\end{equation}
$K^{\rho \mu \nu}$ the "Contorsion Tensor"
\begin{equation}
K^{\rho}_{\,\,\,\,\mu\nu} = \frac{1}{2}g^{\rho\lambda}\bigl(T_{\mu\lambda\nu}+T_{\nu\lambda\mu}+T_{\lambda\mu\nu}\bigr),
\end{equation}
and $T^{\rho \mu \nu}$ the "Torsion Tensor"
\begin{equation}
\label{torscon}
T^\rho_{\,\,\, \mu \nu} = 2 \Gamma^\rho_{\,\,\,  [\mu \nu]}.
\end{equation}
When discarding the curvature and dealing only with torsion, the Christoffel connection can be written in terms of tetrad fields as:
\begin{equation}
\Gamma^{\alpha}_{\,\,\, \mu \nu} = e^\alpha_a \partial_{\mu} e^a_{\nu}
\label{weitz}
\end{equation}
which is called the Weitzenb\"ock connection.

The main properties of TEGR and its applications to cosmology can be found in \cite{Arcos:2005ec, Aldrovandi:2013wha}. Treating gravity as a gauge theory of translation group might solve some ultraviolet inconsistencies, though does not provide any further prediction with respect to GR at the large-scales. However, an extension of the teleparallel action \eqref{tegr action} can be considered into the action through a function of the torsion scalar \cite{Cai}, namely:
\begin{equation}
S=\int d^4x \, |e| \,f(T).
\end{equation} 
By varying the action with respect to the tetrad fields,   second-order field equations arises, that is:
\begin{equation}
\frac{1}{e} \partial_\mu(e \; e^\rho_a S_\rho^{\;\; \mu \nu}) f_T(T) - e^\lambda_a T^\rho_{\;\; \mu \lambda} S_\rho^{\;\; \nu \mu}f_T(T)+ e^\rho_a S_\rho^{\;\; \mu \nu}(\partial_\mu T) f_{TT}(T) + \frac{1}{4} e^\nu_a f(T) = 0
\label{equazioni di campo f(T)}
\end{equation}
Though the teleparallel field equations are exactly equivalent to those of Einstein's GR, $f(R)$ theory and $f(T)$ theory are different. Nevertheless, extended TEGR provides several cosmological predictions not contemplated by GR, as well as modifications to $\Lambda$CDM model \cite{Setare:2013xh}, explanation to the accelerated universe expansion without Dark Energy  \cite{Cai,Sharif:2012ig, Bengochea:2010sg, Bajardi:2021tul} and models for the inflationary paradigm \cite{Ferraro:2006jd}. 

To conclude, the last theory of gravity we considered is the \emph{Horava--Lifshitz} gravity, firstly proposed by Horava in \cite{Horava:2008ih, Horava:2009uw}; it represents a quantum gravity approach aimed to extend GR to the UV scales. According to this theory, the Lorentz invariance is a low-energy symmetry, which breaks at the fundamental level, whose spacetime structure turns out to be anisotropic. Moreover, as pointed out in \cite{Horava:2008jf, Barvinsky:2015kil, Vernieri:2015uma, Vernieri:2011aa, Vernieri:2012ms}, it can provide a solution for the divergences in the two-loop effective action of GR, since it is power counting renormalizable. With a generic line element: 
\begin{equation}
d s^{2}=N^{2} d t^{2}-g_{i j}\left(d x^{i}+N^{i} d t\right)\left(d x^{j}+N^{j} d t\right)
\end{equation}
the Horava-Lifshitz four-dimensional action reads
\begin{equation}
S=\int d^3x \,dt \,  \sqrt{-g} \left\{\frac{2}{\kappa^{2}}\left(K_{i j} K^{i j}-\lambda K^{2}\right) -\frac{\kappa^{2}}{2 w^{4}}\left(\nabla_{i} R_{j k} \nabla^{i} R^{j k}-\nabla_{i} R_{j k} \nabla^{j} R^{i k}-\frac{1}{8} \nabla_{i} R \nabla^{i} R\right)\right\}
\label{actHor}
\end{equation}
with 
\begin{eqnarray}
&& K_{i j}=\frac{1}{2 N}\left(\dot{g}_{i j}-\nabla_{i} N_{j}-\nabla_{j} N_{i}\right)  \qquad K^2 = g_{ij} K^{ij}
\end{eqnarray}
and where $\lambda$, $\omega$ and $\kappa$ are coupling constants. To develop the entropy analysis, in Sec. \ref{HLS} we use the spherically symmetric solution of the above action, provided in \cite{Harko:2009qr}. For further readings see \cite{Nojiri:2010wj, Blas:2009qj, Kiritsis:2009sh, Cai:2009pe} and reference therein. 

The above ones are the theories which we are going to consider in view of the issue of renormalizability of gravitational interaction. 

\section{Spherical Symmetry}
Let us consider now the spherical symmetry for the above theories of gravity to calculate the power counting analysis and the Bekenstein-Hawking entropy.
\label{sectsshog}
\subsection{$f(R)$ gravity}
Let us solve the $f(R)$ field equations in $n$-dimensional spherically symmetric spacetime starting from the  action
\begin{equation}
S = \int \sqrt{|g|} f(R)\; d^n x.
\end{equation} 
The above action can be recast in terms of  Lagrange multiplier $\lambda$ as:
\begin{equation}
S = \int \left[\sqrt{|g|} f(R) - \lambda(R-\tilde{R}) \right] d^n x,
\label{action f(R)}
\end{equation}
being $\tilde{R}$ the spherically symmetric scalar curvature expression. By assuming an interval of the form 
\begin{equation}
ds^2 = e^{\nu(r)} dt^2 - e^{\lambda(r)} dr^2 - r^2 d\Omega_{n-2}^2,
\label{interval}
\end{equation}
the n-dimensional Ricci scalar can be written as \cite{Tangherlini:1963bw, Myers:1986un}:
\begin{equation}
\tilde{R} = e^{-\lambda} \left( \nu'' - \frac{\nu'^2}{2}- \frac{\nu' \lambda'}{2}\right) + \frac{(n-2) e^{- \lambda}}{r}(\nu' - \lambda')- \frac{(n-2)(n-3)}{r^2} (1 - e^{-\lambda}),
\label{n Ricci}
\end{equation}
where the prime denotes the derivative with respect to $r$ and $d\Omega^2_{n-2}$ is the $n-2$ sphere volume element. As a check,  we see that by setting $n=4$, Eq. \eqref{n Ricci} reduces to the well known four dimensional case. By varying  action \eqref{action f(R)} with respect to $R$, we find the Lagrange multiplier $\lambda$:
\begin{equation}
\frac{\delta S}{\delta R} = |g| f_R(R) - \lambda =0 \;\;\;\;\; \to \;\;\; \lambda = |g| f_R(R),
\label{Lagmul}
\end{equation}
being $f_R(R)$ the first derivative of $f(R)$ with respect to the scalar curvature. Replacing Eqs. \eqref{n Ricci} and \eqref{Lagmul} into Eq. \eqref{action f(R)}, the action becomes:
\begin{eqnarray}
S = \int && \left\{e^{\frac{\nu + \lambda}{2}} [f(R) - R f_R(R)] + e^{\frac{\nu - \lambda}{2}} r^{n-2} \left(\nu'' + \frac{\nu'^2}{2}  - \frac{\nu' \lambda'}{2} \right)f_R(R) + \right.\nonumber
\\
&& \left. + (n-2) e^{\frac{\nu - \lambda}{2}} r^{n-3} (\nu' - \lambda') f_R(R) + (n-2)(n-3)e^{\frac{\nu + \lambda}{2}}r^{n-4}(1-e^{-\lambda}) f_R(R) \right\} d^n x.
\end{eqnarray}
By integrating second derivatives and neglecting the surface terms, the point-like Lagrangian finally takes the form:
\begin{equation}
\Lagr = e^{\frac{\nu + \lambda}{2}} [f(R) - R f_R(R)] - (n-2) e^{\frac{\nu - \lambda}{2}} r^{n-3} \lambda' f_R(R) + (n-2)(n-3)e^{\frac{\nu + \lambda}{2}}r^{n-4}(1-e^{-\lambda}) f_R(R) - e^{\frac{\nu - \lambda}{2}} r^{n-2} R' \nu' f_{RR}(R).
\end{equation}
To find out exact solutions of the equations of motion, we focus on the function $f(R) = f_0 R^k$, with $k$ being a  real number parameter. As soon $k=1$ Einstein's GR is recovered. This choice is  motivated by symmetry considerations \cite{Capozziello:2012iea, Capozziello:2007wc} which may be helpful in reducing the dynamics. Specifically, as demonstrated in \cite{Capozziello:2007wc}, a power-law form of $f(R)$ gravity is compatible with the existence of Noether symmetries. The Euler-Lagrange equations provide the following solution in $n$-dimensions:
\begin{equation}
e^\nu = e^{- \lambda} = 1- \frac{2(c_2-c_1n )}{n (n-1)(n-2) } \frac{1}{r^{n-3}} + \frac{ r^2 R_0}{n (n-1) } \qquad k = \frac{n}{2} \qquad R(r) = R_0,
\label{deSittern-D}
\end{equation}
with $c_1$, $c_2$ integration constants. Interestingly, it is worth noticing that the $n$-dimensional solution of the spherically symmetric field equations occurs analytically for $k = n/2$. The only exception is for $k = 1$, where there exists a solution also in four dimensions. As pointed out in the next section, the value $k = n/2$ is the limit for the renormalizability of $R^k$ gravity. Also note that the $n=4$ limit provides
\begin{equation}
e^\nu = e^{- \lambda} =\frac{4 c_1 - c_2 + 12 r + r^3 R_0}{12 r},
\end{equation}
so that $R_0$ plays the role of cosmological constant. Assuming $c_1 = (c_2 - 24 GM)/4$, the solution can be recast as:
\begin{equation}
e^\nu = e^{- \lambda} = 1 - \frac{2 GM}{r} + \frac{R_0}{12} r^2 ,
\end{equation}
which is a Schwarzschild-de Sitter-like solution \cite{Capozziello:2007wc}.

\subsection{ $f(\Gi)$ gravity}
Let us consider now a spherically symmetric background in  $n-$dimensional $f(\Gi)$  gravity. In order to get the point-like Lagrangian, we start from the action
\begin{equation}
S = \int \left[\sqrt{-g} f(\Gi) - \lambda (\Gi - \tilde{\Gi})\right] d^n x;
\end{equation}
with $\tilde{\Gi}$ being the spherically symmetric expression of the Gauss-Bonnet scalar \eqref{GB scalar}, written in terms of the interval \eqref{interval}, namely \cite{Fra&Capo}:
\begin{eqnarray}
\label{GBterm_spherical}
\tilde{\Gi} &=& \frac{e^{-2 \lambda}}{r^4} (n-2)(n-3) \{4 r \lambda' e^{\lambda/2 }[(e^{\lambda}-1)  (n-4) +(e^{ \lambda}-3)e^{\nu/2 } r\nu'] + \nonumber
\\
&+& (e^{ \lambda}-1) [(e^{\lambda}-1) (n-4)(n-5) - 4 (n-4)e^{\nu/2 } r \nu' -4 r^2 e^{\nu/2 }  \nu'^2 - 4 r^2e^{\nu/2 }  \nu''] \}.
\end{eqnarray}
Note that the Gauss-Bonnet term vanishes for $n \le 3$ while turns into a surface term for $n=4$. After varying the action with respect to $\Gi$ and integrating the terms containing higher spatial derivatives, the point-like Lagrangian takes the form:
\begin{eqnarray}
 \label{pointlikeLagra}
\Lagr &=& e^{-2 \lambda} r^{n-6} \{e^{\frac{1}{2}(5 \lambda + \nu)} r^4 (f - \Gi f_\Gi) + (e^{\lambda}-1) (n-2)(n-3) [(n-4)(n-5) e^{(\nu+\lambda)/2} f_\Gi  (e^{\lambda}-1)  + \nonumber
\\
&+&4 e^{3 \lambda/2} r \lambda'] +  4 e^{\lambda/2 + \nu} r^2 \Gi' \nu' f_{\Gi\Gi} \}.
\end{eqnarray}
By choosing a power-law function of the form $f(\Gi) = f_0 \Gi^k$, suggested by symmetry considerations, the solution of the Euler-Lagrange equations turns out to be \cite{Fra&Capo}:
\begin{equation}\label{Pgensol}
e^\nu = e^{-\lambda} = 1 \pm r^{\frac{5-n}{2}}\sqrt{\frac{4 c_1 (n-1)}{120 \binom{n}{n-5} }} \pm r^2 \sqrt{ \frac{ \Gi_0 (n-4) }{120 \binom{n}{n-5} }} \qquad \Gi(r) = \Gi_0 \qquad k = n/4,
\end{equation}
with $c_1, \,  \Gi_0$ constants. Eq. \eqref{Pgensol} is a de Sitter-like solution for $f(\Gi)$ gravity, so that $\sqrt{\Gi_0}$ can be intended as a cosmological constant and $c_1$ as a dimensional coupling with mass dimension $c_1 \sim M^2$. Similarly to  $f(R)$ gravity, the constraint $k=n/4$ can be considered to get analytic solutions. Note that setting $n = 4$, the only possible solution occurs for $k =1$, which does not provide any contribution to  dynamics. Also here, the value $k=n/4$ is the limit value for the renormalizability of $f(\Gi) = \Gi^k$ gravity. As a matter of facts, being constructed by second-order curvature invariants, the renormalizability limit of $\Gi^k$ gravity can be easily inferred by the definition of the Gauss-Bonnet scalar. This link between analytic solutions and power-counting renormalizability of modified theories of gravity represents an interesting feature which deserves to be discussed in details. 

\subsection{ $f(T)$ teleparallel gravity}
 $f(T)$ gravity is another interesting case. Here, the approach is not  straightforward as in the previous extended theories. In fact, the interval yields different sets of tetrad fields which must be selected according to the postulate of "good tetrads" \cite{Tamanini:2012hg}. For this reason, let us focus on the four-dimensional case starting from the action:
\begin{equation}
S = f_0 \int |e| \, T^k \, d^4 x. 
\end{equation}
As a starting point, among all the possible sets of tetrad fields providing the interval \eqref{interval}, it seems reasonable to choose  
\begin{equation}
e^a_\mu = \text{diag}(e^{\frac{\nu}{2}}, \, e^{\frac{\lambda}{2}}, \, r, \, r \sin \theta).
\label{diag tetrads}
\end{equation}
It yields the following torsion scalar \cite{Cai}:
\begin{equation}
\tilde{T} = \frac{e^{-2 \lambda} (1 + r \nu')}{r^2},
\end{equation}
and, after some computations, two classes of solutions are
\begin{equation}
ds^2 = \frac{c_1}{r} dt^2 - \frac{r}{c_2 - 4r} dr^2 - r^2 d\Omega^2
\label{first int}
\end{equation}
and
\begin{equation}
ds^2=\left( \frac{c_4}{c_5 r^8 - 2c_3}\right)^{\frac{1}{8}} dt^2 - \left(\frac{ 2c_3- c_5 r^8}{2r^6} \right) dr^2 - r^2 d \Omega^2.
\label{Tsol}
\end{equation}
Notice that both solutions are independent of the parameter $k$ labeling the power-law function $f(T) = f_0 T^k$. This is due to the fact that diagonal set of tetrads \eqref{diag tetrads} constrains the free parameter $k$ to be $k = \frac{1}{2}$. Therefore, also the function $f(T)$ is further constrained to $f(T) = f_0 \sqrt{T}$. Moreover, solution \eqref{Tsol} can be obtained only after the imposition $ T(r) \sim \text{const}$. Another possibility is to consider a non-diagonal set of tetrads of the form \cite{Cai} 
\begin{equation}
e^a_\mu =
\left(\begin{matrix}
e^{\frac{\nu}{2}} & 0 &0  & 0
\\
0 & e^{\frac{\lambda}{2}} \sin \theta \cos \phi & r \cos \theta \cos \phi  & - r \sin \theta \sin \phi
\\
0 & e^{\frac{\lambda}{2}} \sin \theta \sin \phi & r \cos \theta \sin \phi & r \sin \theta \cos \phi
\\
0 &e^{\frac{\lambda}{2}} \cos \theta & - r \sin \theta& 0
\end{matrix}\right),
\label{non-diag tetrads}
\end{equation}
whose determinant is $|e| = e^{\frac{\nu + \lambda}{2}} r^2 \sin \theta$. Under this assumption, the torsion scalar becomes
\begin{equation}
T= \frac{2 e^{-\lambda}(e^{\frac{\lambda}{2}} - 1)(e^{\frac{\lambda}{2}} - 1 - r \nu')}{r^2}.
\end{equation}
The Lagrange multiplier method allows to find out a suitable  point-like Lagrangian of the form
\begin{equation}
\Lagr = e^{\frac{\nu - \lambda}{2}} \left\{ e^{\lambda} r^2 [f(T) - T f_T(T)]  + 2 f_T(T) (e^{\frac{\lambda}{2}} - 1)(e^{\frac{\lambda}{2}} - 1 - r \nu')\right\}.
\end{equation}
admitting as a possible solution
\begin{equation}
ds^2 = \left(\frac{r}{\ell}\right)^m dt^2 - \frac{1}{n^2} dr^2 - r^2 d\Omega^2,
\label{f(T)sol}
\end{equation}
with $n$ and $m$ constants of integration. Furthermore, a relation among $k$, $n$ and $m$ occurs, \emph{i.e.} 
\begin{equation}
f(T) = f_0 T^k \;\;\;\;\;\; k= \frac{m^2 + 4(n-1)^2}{4(2+m-2n)} \neq 0.
\end{equation}
This solution is directly linked to the \emph{ansatz}  in \eqref{non-diag tetrads}, while other teleparallel black hole solutions can be found in \cite{Ferraro:2011ks, Wang:2011xf, HamaniDaouda:2011iy, Atazadeh:2012am, Nashed:2014sea, Aftergood:2014wla, Nashed:2011fz, Nashed:2015wia}. 

\subsection{Horava-Lifshitz gravity}
\label{HLS}
The last alternative theory of gravity we are going to  discuss is the four-dimensional Horava-Lifshitz theory. Let us take into account   action \eqref{actHor} in the spherically symmetric background \eqref{interval} with $n = 2$. This choice, as pointed out in \cite{Harko:2009qr} and \cite{Kehagias:2009is}, yields the solution:
\begin{equation}
e^\nu = e^{- \lambda} = 1+\left(\omega-\Lambda \right) r^{2}-\sqrt{r\left[\omega\left(\omega-2 \Lambda \right) r^{3}+\beta\right]},
\end{equation}
with $\beta$ and $\omega$ constants. Equivalently, by means of the assumption $\beta = 4 \omega G M$ and $\Lambda = 0$, the solution can be written as
\begin{equation}
e^{\nu}=1+\omega r^{2}-\omega r^{2} \sqrt{1+\frac{4G M}{\omega r^{3}}},
\end{equation}
which, in the second-order expansion  $\displaystyle \frac{4 GM}{\omega r^{3}} \ll 1$, provides a Schwarzschild-like interval.
\section{Power Counting Analysis}
\label{sectpca}
The renormalizability of the above theories of gravity can be developed according to a power-counting analysis.  As a first remark of this section, we point out that the analysis of the coupling constant mass dimension can provide only an indication about the renormalizability. Indeed, most of the above treated modified theories of gravity might be not-renormalizable when more detailed aspects are considered, as well the propagator homogeneity in the momenta space \cite{Asorey:1996hz}. Detailed discussions on the renormalizability of $f(R)$ theories of gravity can be found \emph{e.g.} in \cite{Steinwachs:2020jkj, Ruf:2017bqx, Steinwachs:2011zs, Ruf:2017xon}. One of the first attempts towards a renormalization scheme was pursued by Stelle in \cite{Stelle:1976gc}. He showed that the gravitational action can be renormalized only when all possible combinations of second-order curvature invariants are considered along with the Hilbert-Einstein action. Adding terms with more than four derivatives, indeed, makes the theory finite after a certain order in the loop expansion, but does not remove the divergences at the one-loop order.

However, in this case, we follow the same prescription as \cite{Shomer:2007vq} with the aim to find inconsistencies between the two approaches. Moreover, power counting can suggest possible candidates for a perturbatively renormalizable theory. 

We first consider the Einstein--Hilbert action and its $f(R)$ extension, which is a good starting point for dealing with theories constructed by the scalar curvature and its derivatives. Then, we find the renormalizability limit of the modified Gauss--Bonnet gravity and, finally, we apply the same approach to teleparallel actions. For this purpose, the actions must be rewritten in terms of a sum between a free term, acting as a Gaussian fixed point, and other higher perturbative contributions. In this way, it is possible to check the mass dimension of the coupling constant of each term. The first step is to recast the metric tensor as a perturbation of the Minkowski background, {i.e.},
\begin{equation}
g_{\mu \nu} = \eta_{\mu \nu} + h_{\mu \nu},
\end{equation}
so  the Ricci scalar can be written in terms of $\eta_{\mu \nu}$ and $h_{\mu \nu}$. Considering the Levi-Civita connection
\begin{equation}
\Gamma^\rho_{\mu \nu} = \frac{1}{2} (\eta^{\rho \lambda} - h^{\rho \lambda}) ( \partial_\nu h_{\lambda \mu} + \partial_\mu h_{\nu \lambda} - \partial_\lambda h_{\mu \nu})
\end{equation}
and the form of the scalar curvature
\begin{equation}
R = (\eta^{\mu \nu} - h^{\mu \nu})(\nabla_\rho \Gamma^\rho_{\mu \nu} - \nabla_\nu \Gamma^\rho_{\mu \rho}),
\end{equation}
it turns out that the perturbed Ricci scalar depends on second derivatives of $h_{\mu \nu}$ multiplied by $h_{\mu \nu}$ itself. Since we are not interested in writing the exact form of $R$, rather to check the mass dimension of each term, a naive procedure permits to recast the Einstein-Hilbert action as:
\begin{equation}
S = \frac{1}{\chi^2} \int \left[(\partial \partial h) + (\partial h \partial h) + h (\partial h \partial h) + h^2 (\partial h \partial h) + ... \right] d^4x,
\label{GR PW action}
\end{equation} 
with $\chi$ being a generic coupling constant. Notice that the notation does not take into account constant factors, while it only considers the amount of $h$ and $\partial h$ occurring in each term. For instance, the quantity $\partial h \partial h$ includes all terms made of the product of first derivatives of the tensor $h_{\mu \nu}$. 
By an appropriate gauge choice,  it is always possible to delete the first term. Thus, by means of the reparameterization $h = \chi \tilde{h}$, the action can be written as:
\begin{equation}
S = \int \left[ (\partial \tilde{h} \partial \tilde{h}) + \chi \tilde{h} (\partial \tilde{h} \partial \tilde{h}) + \chi^2 \tilde{h}^2 (\partial \tilde{h} \partial \tilde{h}) + ... \right] d^4x.
\label{LFR}
\end{equation}
In this form, the action accounts for a perturbation of the Gaussian fixed point $\int (\partial\tilde{h})^2 d^n x$, which is the expression needed in order to check the coupling constant mass dimension of any terms. To this aim, there is no need to study the mass dimension of higher-order terms, since it can be inferred from the mass dimension of the first one. Because of the negative mass dimension of $\chi$, the above action shows the reason why GR is not a renormalizable field theory.

Considering the linearized form of $R$ in Eq. \eqref{LFR}, the prescription can be easily extended to the more general action
\begin{equation}
S = \frac{1}{\chi^2} \int \sqrt{|g|} R^k \; d^nx.
\label{initial action f(R)}
\end{equation} 
Now, by the expansion \eqref{GR PW action}, the action \eqref{initial action f(R)} (up to first order) reads as:
\begin{eqnarray}
S &=& \frac{1}{\chi^2} \int \left[(\partial h \partial h) + h (\partial h \partial h)\right]^k d^n x = \frac{1}{\chi^2} \int  \sum_{i=0}^k \binom{k}{i} (\partial h \partial h)^{k-i} h^i (\partial h \partial h)^i \, d^n x = \frac{1}{\chi^2} \int  \sum_{i=0}^k \binom{k}{i} (\partial h \partial h)^k h^i \, d^n x = \nonumber
\\
&=& \frac{1}{\chi^2} \int \left[(\partial h \partial h)^k + k h  (\partial h \partial h)^k   + \frac{k\left(k-1 \right)}{2} h^2  (\partial h \partial h)^k +... \right] \, d^n x =  \nonumber
\\
&=& \frac{1}{\chi^2} \int \left[(\partial h \partial h)^k \left(1 + k h  + \frac{k\left(k-1 \right)}{2} h^2 + ... +  h^k \right) \right] \, d^n x \sim \frac{1}{\chi^2} \int \left[(\partial h \partial h)^k + k h (\partial h \partial h)^k \right] \, d^n x.
\end{eqnarray}
Similarly to the previous case, to check the coupling constant dimension,  we can introduce the definition $h = (\chi)^{\frac{1}{k}}\tilde{h}$, so that the action takes the form
\begin{equation}
S = \int \left[(\partial \tilde{h} \partial \tilde{h})^k + k \, \chi^{\frac{1}{k}}\tilde{h} (\partial \tilde{h} \partial \tilde{h})^k \right] \, d^n x.
\end{equation}
From the initial action \eqref{initial action f(R)},  it is possible to infer the mass dimension of $\chi$ in $n$-dimensional $R^k$ gravity, that is $\displaystyle \chi \to [M^{k-\frac{n}{2}}]$. Therefore power-law  $f(R)$  gravity, with $f(R)$ must satisfy at least one of the following conditions
\begin{equation}
\label{pcfr}
1- \frac{n}{2k} \ge 0 \;\;\;\; \to \;\;\;\; k< 0  \; \lor \; k \ge \frac{n}{2}.
\end{equation}
to be power-counting renormalizable. 
Thus, according to the power counting argument, only $f(R) = R^2$ gravity is renormalizable in four dimensions. In addiction, all  power-law models of the form $f(R) =R^k$, with $k>2$, are super-renormalizable. 

By similar reasons, it is straightforward to show that higher-order theories with action $ S = \int \sqrt{|g|} \Box^{\ell} R^k \; d^nx$ must satisfy the conditions
\begin{equation}
\begin{cases}
& \displaystyle k \ge \frac{n}{2} - \ell \; \lor \; k \le 0 \;\;\; \text{if} \; n> 2 \ell,
\\
\\
& \displaystyle k \le \frac{n}{2} - \ell \; \lor \; k \ge 0 \;\;\; \text{if} \; n< 2 \ell,
\end{cases}
\end{equation} 
to be at least renormalizable. Notice that this result holds for any $\ell \in \mathbb{Z}$, which means that even negative powers of the operator $\Box$ can occur in the action. In this case, the corresponding theory turns out to be non-local \cite{Capriolo1,Capriolo2}. The most general higher-order action containing one single term which depends on the d'Alembert operator and on the scalar curvature is
\begin{equation}
S = \frac{1}{\chi^2} \int \sqrt{|g|} R^k \Box^\ell R^p \; d^n x.
\label{action0}
\end{equation}
In this case, the mass dimension of the coupling constant is $[\chi] = M^{k+\ell + p - \frac{n}{2}}$, and the first-order linearization of the action \eqref{action0} reads as
\begin{equation}
S = \frac{1}{\chi^2} \int \left[(\partial h \partial h)^k + h (\partial h \partial h)^k \right]\left\{\Box^\ell (\partial h \partial h)^p +\Box^\ell  \left[h (\partial h \partial h)^p \right] \right\} d^nx.
\label{act123}
\end{equation}
The reparameterization of $h$, which allows to write the action in the form required by power-counting analysis,  is $h = \tilde{h} \chi^{\frac{1}{p+k}}$. Therefore Eq. \eqref{act123} becomes:
\begin{equation}
S = \int (\partial \tilde{h} \partial \tilde{h})^k \Box^\ell (\partial \tilde{h} \partial \tilde{h})^p \, d^n x + \int \left\{ (\partial \tilde{h} \partial \tilde{h})^k \Box^\ell \left[\tilde{h} (\partial \tilde{h} \partial \tilde{h})^p\right] + \tilde{h} (\partial \tilde{h} \partial \tilde{h})^k \Box^\ell  (\partial \tilde{h} \partial \tilde{h})^p \right\} \chi^{\frac{1}{p+k}} \,d^nx.
\end{equation}
The conditions which makes the theory power-counting renormalizable are therefore
\begin{equation}
\begin{cases}
&\displaystyle k \ge \frac{n}{2} - p - \ell \; \lor k \le -p \;\;\;\; \text{if} \; n> 2\ell,
\\
\\
&\displaystyle k \le \frac{n}{2} - p - \ell \; \lor k \ge -p \;\;\;\; \text{if} \; n< 2\ell.
\end{cases}
\label{non local ricci}
\end{equation}
This general result can be extended to actions containing higher order curvature invariants. To this purpose, we consider as equivalent all those scalar quantities with same mass dimensions. This is the case of the Gauss-Bonnet term $\Gi$ and the quadratic scalar curvature $R^2$. In the modified Gauss-Bonnet gravity with action
\begin{equation}
S = \frac{1}{\chi^2} \int \sqrt{|g|} \Gi^k \Box^\ell \Gi^p \; d^n x,
\end{equation}
the relation \eqref{non local ricci} can be recast as: 
\begin{equation}
\begin{cases}
&\displaystyle k \ge \frac{n}{4} - p - \ell \; \lor k \le -p \;\;\;\; \text{if} \; n> 4\ell,
\\
\\
&\displaystyle k \le \frac{n}{4} - p - \ell \; \lor k \ge -p \;\;\;\; \text{if} \; n< 4\ell.
\end{cases}
\label{GB}
\end{equation}
Specifically, in the limit $\ell = p = 0$, the condition $k> \frac{n}{4}$ is recovered. As a consequence, the four-dimensional action
\begin{equation}
S =\frac{1}{\chi^2} \int \sqrt{-g} \Gi^k \, d^4 x ,
\end{equation}
is not-renormalizable only in the range  $0<k<1$. 

To conclude, the last example is given by the action 
\begin{equation}
S =\frac{1}{\chi^2} \int \sqrt{-g} R^k \Gi^m \; d^n x,
\end{equation}
considered \emph{e.g.} in \cite{Benetti:2018zhv, Capozziello:2016eaz}. The resulting power counting analysis can be inferred from the previous cases, so that the theory turns out to be (super-)renormalizable if:
\begin{equation}
k \ge \frac{n}{4} - m \; \lor k\le -m 
\end{equation}
It is worth pointing out the analogy with the teleparallel case, where the linearization of the metric is replaced by the expansion of the tetrad fields around the unity matrix, namely:
\begin{equation}
e^a_\mu = \delta^a_\mu + h^a_\mu.
\label{eamu}
\end{equation}
By means of Eqs. \eqref{eamu} and \eqref{weitz}, the first-order teleparallel action can be written as:
\begin{equation}
S = \frac{1}{\chi^2} \int \left( \partial h \partial h +  h \partial h \partial h \right) d^n x,
\end{equation}
which is formally equivalent to the first-order Einstein-Hilbert action. This implies that the same results, provided in Eq. \eqref{non local ricci}, hold also in this case and a separate treatment is not needed. This is mainly due to the mass dimension of the torsion scalar $T$, which turns out to be the same as that of the scalar curvature.

\section{Renormalizability via Bekenstein-Hawking Entropy Argument}
\label{sectrvbhea}
The renormalizability  of theories of  gravity can be studied also  by comparing their asymptotic behavior to that of a CFT in the appropriate dimension. To introduce the approach, let us first treat the case of $n-$dimensional GR with cosmological constant, whose spherically symmetric  solution reads
\begin{equation}
ds^2 = \left(1- \frac{\eta_n G M}{r^{n-3}} - \frac{\Lambda}{3} r^2\right)dt^2  - \frac{1}{ 1- \frac{\eta_n G M}{r^{n-3}} - \frac{\Lambda}{3} r^2}dr^2 - r^2 d\Omega_{n-2}^2,
\label{n-d Schw}
\end{equation}
being $\eta_n$ a constant depending on the spacetime dimension. The Bekenstein-Hawking formula $\Sw = \frac{A}{4G}$ can be used to infer the relation between $\Sw$ and the energy $\E$. According to the entropy argument, if such dependence is the same as that of a $n$-dimensional CFT, the theory turns out to be renormalizable. For a CFT the energy turns out to scale as
\begin{equation}
\frac{\Sw}{\E} \sim \frac{(RT)^{n-1}}{R^{n-1} T^n} \;\;\; \to \;\;\; \Sw \sim \E^{\frac{n-1}{n}}.
\label{CFT entropy}
\end{equation}
so that, in four dimensions, the condition $\Sw \sim \E^{\frac{3}{4}}$ is satisfied. To compare Eq. \eqref{CFT entropy} with the entropy provided by $n$-dimensional Schwarzschild black holes, we must distinguish the cases $\Lambda r^2 \ll 1$ and $\Lambda r^2 \gg 1$. In the former case, the time-like component simply becomes
\begin{equation}
e^\nu = 1- \frac{\eta_n G M}{r^{n-3}},
\end{equation}
so that the horizon sits at $\tilde{r} \sim GM^{\frac{1}{n-3}}$. Therefore, the Bekenstein-Hawking entropy has the following trend:
\begin{equation}
\Sw = \frac{A}{4G} \sim \frac{r^{n-2}}{G} \sim \E^{\frac{n-2}{n-3}}.
\label{SCHW entropy}
\end{equation}
By comparing Eq. \eqref{SCHW entropy} with Eq. \eqref{CFT entropy} it turns out that there are no dimensions in which GR can be renormalized. On the other hand, in the latter case ($\Lambda r^2 \gg 1$) no terms can be neglected in the time-like component of the interval \eqref{n-d Schw} and the solution can be written as:
\begin{equation}
1- \frac{\eta_n G M}{r^{n-3}} - \frac{\Lambda}{3} r^2 = \frac{r^{n-3} - \eta_n G M - r^{n-1} \frac{\Lambda}{3}}{r^{n-3}} =\frac{\Lambda}{3}\left(\frac{ 3 \frac{r^{n-1}}{r^2 \Lambda} - 3 \frac{\eta_n G M}{\Lambda} - r^{n-1}}{r^{n-3}}\right).
\end{equation}
Considering that,  for the first term the condition $\displaystyle \frac{r^{n-1}}{r^2 \Lambda} \ll 1$ must hold, similar computations as before show that the horizon is:
\begin{equation}
\tilde{r} \sim \left(\frac{G M}{\Lambda} \right)^{\frac{1}{n-1}}
\label{horizon}
\end{equation}
and the entropy scales as
\begin{equation}
\Sw \sim \frac{r^{n-2}}{G} \sim \E^{\frac{n-2}{n-1}}.
\end{equation}
Also here, according to the entropy argument, the Schwarzschild-de Sitter theory cannot be renormalized in any dimensions, in agreement with the power counting analysis in Sec. \ref{sectpca}. However, as it will be showed in the next section, this represents the only case in which the two approaches agree.

\subsection{The $f(R)$ Case}

Let us now follow a similar procedure for spherically symmetric solutions coming from GR extensions, starting from  $f(R)$ gravity. As showed in the previous section, $f(R)$ theory admits the following field equations solution:
\begin{equation}
e^\nu = e^{- \lambda} = \frac{c_2 (n-4) r^{3-n} - 24 G M n r^{3-n} + 2 (n-2) (n^2 -n +r^2 R_0)]}{2 n (n-1)(n-2) } \qquad k = \frac{n}{2},
\end{equation}
where the constant $c_1$ can be identified with $GM$. By setting $c_2 = 0$, the solution becomes:
\begin{equation}
 e^\nu = e^{- \lambda} = \frac{r^{3-n} \left(-12 \frac{G Mn}{R_0} +  \frac{n(n-1)(n-2)}{R_0} r^{n - 3} + r^{n-1}\right)}{2 n (n-1)(n-2) } \qquad k = \frac{n}{2}
\end{equation}
and, under the assumption $ \displaystyle \frac{n(n-1)(n-2)}{R_0} r^{n - 3} \ll 1$, the horizon sits at:
\begin{equation}
\tilde{r} \sim M^{\frac{1}{n-1}}.
\end{equation}
By applying the Bekenstein-Hawking formula and comparing the energy scale with Eq. \eqref{SCHW entropy}, we notice that the entropy argument imposes the theory to be non-renormalizable regardless of the value of $n$. This result disagrees with the power counting analysis in Sec. \ref{sectpca}. Furthermore, noticing that the solution of the $n-$dimensional modified $f(R)$ gravity can be recast as an Einstein-de Sitter solution, the horizon must be the same as Eq. \eqref{horizon}. 
\subsection{The $f(\Gi)$ Case}
With the purpose to study the Bekenstein-Hawking entropy provided by $f(\Gi) = f_0 \Gi^k$ gravity, let us rewrite the spherically symmetric solution \eqref{Pgensol} as:
\begin{equation}
e^{\nu} = \frac{r^{\frac{n-5}{2}} \pm \eta_{1,n} M  \pm r^{\frac{n-1}{2}} \eta_{2, n}}{r^{\frac{n-5}{2}}},
\end{equation}
with $\eta_{1,n} $ and $\eta_{2,n}$ being constants depending on the dimension. In the limit $r^{\frac{n-5}{2}} \ll r^{\frac{n-1}{2}} \eta_{2, n}$ the horizon and the entropy scale as
\begin{equation}
\tilde{r} \sim M^{\frac{2}{n-1}} \qquad \Sw = \frac{A}{4G} \sim \frac{r^{n-2}}{G} \sim \E^{\frac{2(n-2)}{n-1}}.
\end{equation}
This means that the entropy argument provides no values of $k$ and $n$ for which $f(\Gi) =f_0 \Gi^k$ gravity is renormalizable. On the other hand, the analysis in Sec. \ref{sectpca}, suggests that the theory is power-counting renormalizable for any $k \ge n/4$, at odds with the entropy analysis. 

\subsection{The Teleparallel Case}
In the teleparallel case, the power-law solution \eqref{f(T)sol} has  no event horizon besides $r=0$, so that the comparison between the two approach cannot be applied. Let us therefore consider solution \eqref{Tsol}, coming from a diagonal set of tetrad fields. In this case, the torsion scalar is constant: $T = T_0$. Due to the general structure of the theory, some difficulties arise in recasting the ratio $\displaystyle \frac{c_3}{c_5}$ as a mass term. This is due to the presence of torsion to label the spacetime, which implies that the spin of the compact object must be somewhere considered. However, following Ref. \cite{Cai} and identifying the constant $c_5$ with $T_0$, the horizon sits at:
\begin{equation}
\tilde{r} \sim T_0^\frac{1}{8}.
\end{equation}
Since the energy is linearly proportional to the torsion, the entropy scales as:
\begin{equation}
\Sw \sim \frac{A}{4G} \sim \E^{\frac{1}{4}},
\end{equation}
and suggests that the modified teleparallel gravity with function $f(T) = f_0T^k$ is not renormalizable for any $k$. On the contrary, according to the power-counting renormalization scheme, the function $f(T) =f_0 T^k$ is renormalizable for any $k > 2$. Also here the two approach are not in agreement.

\subsection{The Horava-Lifshitz Case}
To conclude, we show that also the analysis of Horava-Lifshitz gravity provides disagreements between the entropy and the power counting arguments. Though we have not considered Horava-Lifshitz gravity in Sec. \eqref{sectpca}, we take for granted the result provided in Ref. \cite{Horava:2008jf}, where the author shows that the theory is power counting renormalizable. 

Considering the spherically symmetric solution provided in \cite{Harko:2009qr}, the horizon is:
\begin{equation}
\tilde{r}=GM[1 - \sqrt{1-1 /\left(2 \omega GM^{2}\right)}].
\end{equation}
In the limit $2 \omega GM^2 \gg 1$, the horizon turns out to have the same form as in GR, namely $\tilde{r} \sim 2GM$, so that the theory can be considered non-renormalizable according to the entropy argument. In general, without assuming any further constraints, the entropy $\mathcal{S} \sim \tilde{r}^{n-2}$ behaves differently than that of  CFT, denying the validity of this method to test the renormalizability of alternative theories of gravity.

\section{Conclusions}
In this paper,  we addressed some concerns claiming for  "no-go" arguments to the asymptotic safety of GR, also taking into account power counting analysis of modified theories of gravity as  counterexamples. In particular, we addressed the analysis to the so-called entropy argument \cite{Shomer:2007vq}, which is naturally based on physical intuition. However, the intuition behind it seems unjustified when applied to energy scales far beyond the scales of the theories from which they come from, as pointed out in Ref. \cite{Doboszewski2018}. One of the main premise of the entropy argument is the presupposition of the validity of the Bekenstein-Hawking entropy formula at high energy scales, but there is no \emph{a priori} reason to expect the Bekenstein-Hawking entropy to be a well-defined quantity at Planck scales. Furthermore, assuming the Bekenstein-Hawking formula, working at high energies, means that its validity is exact, but quantum corrections to the formula, as showed \emph{e.g.} in \cite{Sen2013, Bytsenko2013}, could be a strong indication that this assumption is not justified. Another assumption on which the entropy argument is based on is that of black hole dominance at high energies, extrapolated from the classical theory. Unfortunately, so far, very little knowledge concerns mini  and quantum black holes.

In summary, we briefly introduced some  modified theories of gravity and wrote down the related explicit spherically symmetric solutions. Then, we performed a naive power counting analysis, which at least in  these cases, can give   sufficient conditions. This was done in order to make a comparison with the entropy argument. What we derived is that, even for these simple cases, the two methods disagree. Moreover, as pointed out throughout the paper, $f(R) = R^k$ field equations can be analytically solved in a $n-$dimensional spherically symmetric background only when the relation $ k = n/2$ occurs. This result is even more interesting if compared to $f(\Gi)$ gravity. Indeed, it turns out that, in the same background, $f(\Gi) = \Gi^k$ gravity admits analytic solutions only when $k = n/4$. 

Of course, this is not a proof of the overall validity of the entropy argument, which has to be better investigated in modified gravity, as we will do in future papers. However, the approach is in agreement with the results provided by Ref. \cite{Falls:2012nd}, according to which relation of entropy density versus energy density must be adopted to check renormalizability. 
 
We also aim to study the power counting renormalizability and unitarity of other modified theories of gravity, such as $R^2 + R_{\mu \nu} R^{\mu \nu}$, which is scale-invariant and multiplicatively renormalizable, but may have some problems with unitarity.

\section*{Acknowledgement}
The Authors acknowledge the support of {\it Istituto Nazionale di Fisica Nucleare} (INFN) ({\it iniziative specifiche} GINGER, MOONLIGHT2, and QGSKY).

\end{document}